\documentclass[
    ,final            
  ]
  {aipproc}

\layoutstyle{6x9}

\begin{document}

\title{Novae In External Galaxies From The POINT-AGAPE Survey And The Liverpool
Telescope}

\author{M. J. Darnley}{
  address={Astrophysics Research Institute, Liverpool John Moores University,
Birkenhead, CH41 1LD, UK}
}

\author{M. F. Bode}{
  address={Astrophysics Research Institute, Liverpool John Moores University,
Birkenhead, CH41 1LD, UK}
}

\author{E. J. Kerins}{
  address={Astrophysics Research Institute, Liverpool John Moores University,
Birkenhead, CH41 1LD, UK}
}

\author{T. J. O'Brien}{
  address={Jodrell Bank Observatory, The University of Manchester,
Macclesfield, SK11 9DL, UK}
}

\begin{abstract}
We have recently begun a search for Classical Novae in M31 using three
years of multicolour data taken by the POINT-AGAPE microlensing
collaboration with the 2.5m Isaac Newton Telescope (INT) on La~Palma.
This is a pilot program leading to the use of the Liverpool Telescope
(LT) to systematically search for and follow novae of all speed
classes in external galaxies to distances up to around 5Mpc.
\end{abstract}

\maketitle


\section{Introduction}

The importance of the study of novae in external galaxies has been recognised
since the time of Hubble \cite{Hubble:1929}.  However, there are great
difficulties in obtaining the frequency and duration of observations with
large enough telescopes to determine the peak magnitude and speed class for a
meaningful sample of objects (see papers by Shafter, Shara and Della Valle,
this volume).

We aim to search for and follow the temporal development of novae
across a range of speed classes in selected galaxies over several
years.  Our goals are to tighten the MMRD, $t_{15}$ and other
relationships, which make classical novae potentially very important
distance indicators \cite{Warner:NovaBook}.  We hope to help
resolve the debate about the dependence of nova rate with galaxy type
and stellar population \cite{DellaValle:1994,Yungelson:1997}. Finally, we
wish to determine whether there are indeed two distinct populations of
classical novae \cite{DellaValle:1992} and how this might relate to current
models of the outburst.

This project is being pushed on two fronts: first with existing data
from the POINT-AGAPE gravitational microlensing survey and then with
the soon-to-be-commissioned Liverpool Telescope (LT).  We plan to automate
as far as possible the data reduction and novae selection
processes.

\subsection{Identification of Novae}

We are identifying novae using a method similar to that outlined in Shafter et
al \cite{Shafter:2000}.  The images are first carefully aligned and then
have their point-spread functions matched.  We identify strongly varying
sources by subtracting the aligned and matched images from one another and
searching for any residual sources.  Photometry is carried out on
the data using the DAOPHOT package within the IRAF environment.

In this dataset we are sensitive to objects down to $r' \simeq 23$,
though our sensitivity is a strong function of position and decreases towards
the bulge.  Ultimately we plan to use simulated datasets to correct for this
spatial bias.

\section{The POINT-AGAPE Survey}

\begin{figure}
  \includegraphics[height=.4\textheight]{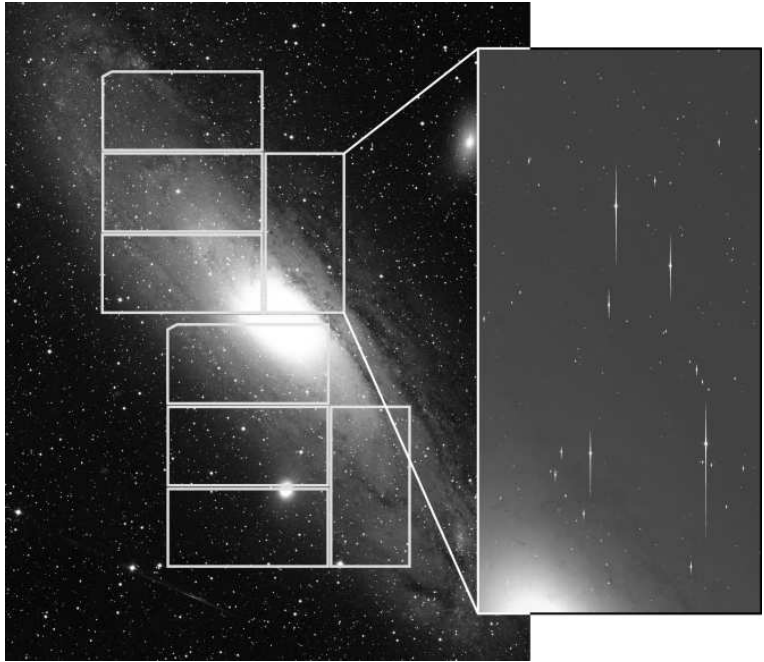}
  \caption{The Andromeda Galaxy and the positions of the two POINT-AGAPE
fields.  (Image adapted from an original image obtained by Bill Schoening,
Vanessa Harvey/REU programme/AURA/NOAO/NSF.)  Inset shows one of
the four chips from one of our INT WFC frames.}
  \label{pa}
\end{figure}

POINT-AGAPE is an Anglo-French collaboration, which is searching for
gravitational microlensing events against the mostly unresolved stars
in M31 \cite{ker01}.  The survey team is employing a modified
detection technique dubbed "pixel lensing" developed to cope with the
problems associated with temporal variations in seeing and sky
background.  The project has evolved from the pilot AGAPE (Andromeda
Galaxy Amplified Pixels Experiment) programme conducted at Pic du
Midi, which probed the M31 bulge \cite{ans97}.  POINT-AGAPE is using
the wide-field camera (WFC) on the INT to survey a much larger area
(POINT is an acronym for Pixel-lensing Observations with the INT).
Since autumn 1999, two fields covering 0.6 square degrees of the M31
disk have been monitored regularly in three colours (Sloan-like $g'$, $r'$
and $i'$ filters, which correspond to wavelengths of 486, 622 and 767
nm, respectively).  The locations of the two fields are shown in
Figure~\ref{pa}.  To date three seasons of data comprising roughly 180 epochs
have been collected.

The prerequisites for the detection of microlensing are good temporal sampling
and a long baseline of observations and it is these same qualities that make
the POINT-AGAPE dataset an excellent repository for novae searches.  Since
novae are much brighter than typical microlensing events they should, for the
most part, show up as resolved sources, negating the need for us to employ the
pixel-lensing method.

However, with around 50GB of data to analyse, the task of finding and
cataloging novae in M31 remains a challenging one. To meet this
challenge we are developing algorithms to automatically reduce the
data and search for novae.  Parts of this software have already
allowed us to discover two probable novae candidates in a small subset
of the dataset provided by the POINT-AGAPE collaboration.

\subsection{First Results}

\begin{figure}
  \includegraphics[width=1\textwidth]{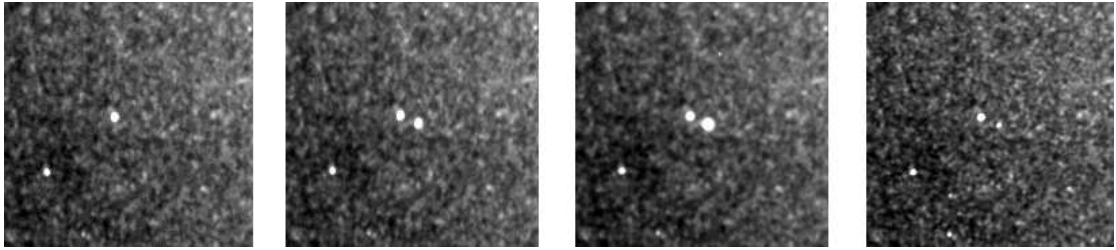}
  \caption{An $r'$-band image sequence of the first nova candidate from the
  POINT-AGAPE INT survey.}
  \label{nova}
\end{figure}

\begin{figure}
  \rotatebox{270}{\includegraphics[height=.6\textheight]{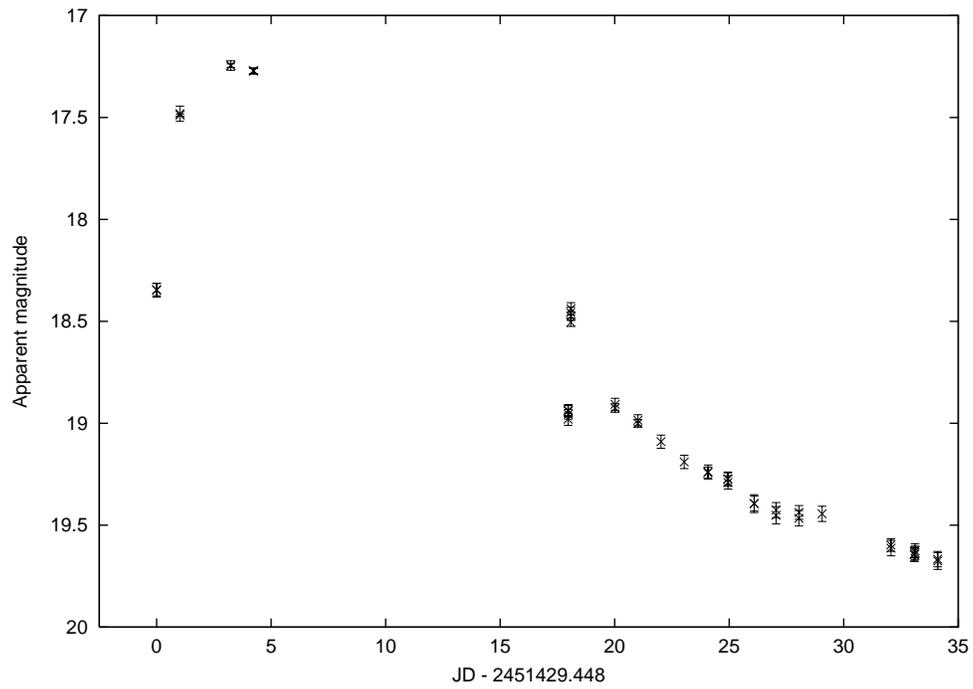}}
  \caption{$r'$-band lightcurve of the first nova candidate from the
  POINT-AGAPE survey data.}
  \label{lightcurve1}
\end{figure}

\begin{figure}
  \rotatebox{270}{\includegraphics[height=.6\textheight]{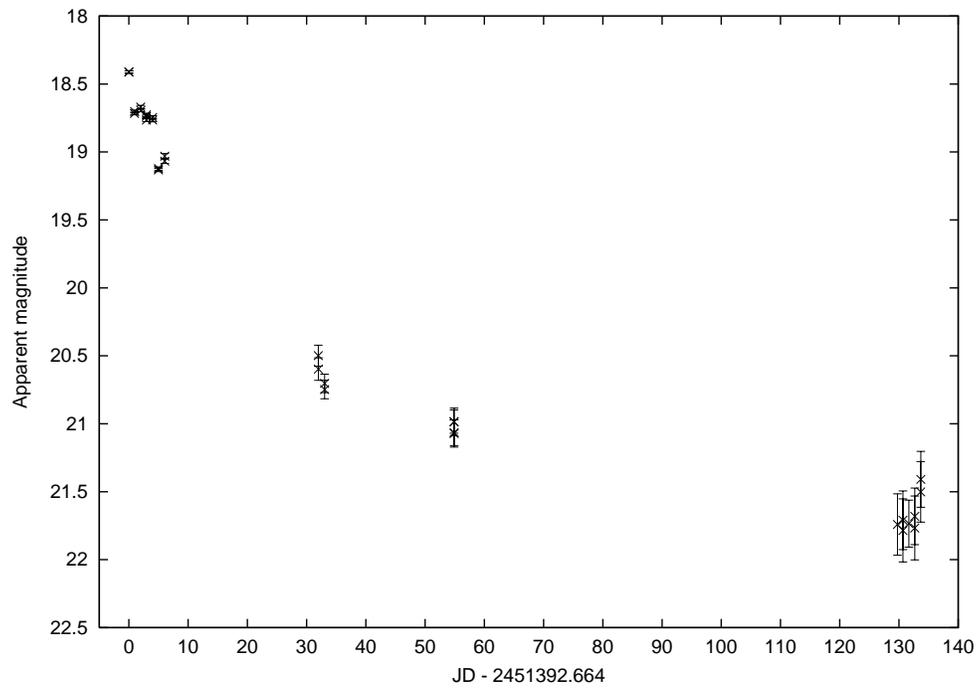}}
  \caption{$i'$-band lightcurve of the second nova candidate from the
  POINT-AGAPE survey data.}
  \label{lightcurve2}
\end{figure}

We have identified two nova candidates in the POINT-AGAPE
dataset. Both novae were initially discovered in the $r'$-band data
and then located in both the $g'$ and $i'$ bands.  The first nova
candidate became visible on 7th September 1999, was caught during the
initial rise in both $r'$ and $g'$ bands, and was subsequently followed to
provide an estimate of $t_2 \simeq 20$~days, classing this candidate as a
fast nova \cite{Payne:1957}.  Figure~\ref{nova} shows a sequence of $r'$-band
images which straddle the outburst. The second nova was discovered near
maximum at the beginning of the first observing season on 2nd August
1999. We estimate $t_2 \simeq 30$~days for this event, making it a moderately
fast nova \cite{Payne:1957}.
Figures~\ref{lightcurve1} and \ref{lightcurve2} show the $r'$-band and
$i'$-band lightcurves for the first and second nova
candidates, respectively.  The photometric calibration of these
lightcurves is still being worked on, but these results illustrate the
types of nova observations that we will be able to secure.

The two novae candidates have been found whilst testing our detection
methods on a small 1024x1024 pixel section of the data over a period
of about three months, amounting to less than 0.3\% of the entire
dataset.  We are now beginning to search for variable objects in the
whole dataset and, based on estimates of the M31 novae rate of $29 \pm
4$~yr$^{-1}$ \cite{Capaccioli:1989}, we are expecting to find 30-50
classical novae.  One of the advantages of our
dataset is excellent temporal coverage which should allow us
to produce reasonably complete lightcurves, hopefully with many sampling the
all-important time of maximum light.

\section{The Liverpool Telescope}

The Liverpool Telescope, with a primary mirror diameter of 2m, will be the
World's largest fully-robotic telescope.  The LT is currently being reassembled
at the Observatorio del Roque de Los Muchachos, La~Palma.  Its primary
instruments are a CCD camera with a 12-position filter wheel, a near-IR camera
and a fibre-fed spectrograph.  The telescope is able to move rapidly to
acquire targets and will have arc-second pointing.

The main scientific goals of the LT are: (i) the monitoring of
variable objects on all timescales, from seconds to years; (ii)
the ability to rapidly react to unpredictable phenomena and their systematic
follow up; (iii) simultaneous or coordinated observations with other
facilities, both on the ground and in space; and (iv) small-scale surveys
and serendipitous source follow-up.  Further details about the LT can
be found at the telescope's website
(\url{http://telescope.livjm.ac.uk}).

Guaranteed time has been approved for our novae in external galaxies
program and the telescope is expected to see first light in the autumn
of 2002.  We will use the unique capabilities of the LT to
systematically monitor a carefully selected set of external galaxies;
M81, M64 and M94, to discover and follow as complete a sample of novae
as possible in each system.  We shall be able to detect objects down
to $m_{V} \approx 24$.  Over a three-year programme, we will discover
more classical novae than have been discovered in our own Galaxy to
date, and in many cases provide better lightcurve and distance
determinations.  A by-product of the survey will be the discovery of
many other luminous variables in our target galaxies.

\section{Acknowledgments}
We wish to thank the POINT-AGAPE collaboration for use of its
data. MJD would like to thank Andy Newsam for help with preparing some of the
figures. MJD's research is supported by a PPARC PhD studentship.

{}

\end{document}